\documentclass[12pt]{iopart}
\usepackage{amssymb}
\usepackage{graphicx}
\DeclareMathAlphabet{\mathpzc}{OT1}{pzc}{m}{it}

\usepackage{iopams}
\begin{document}

\title[]{Spatial confinement effects on quantum harmonic oscillator I: Nonlinear coherent state approach}

\author{M. Bagheri Harouni, R. Roknizadeh, M. H. Naderi}

\address{Quantum Optics Group, Department of physics, University of Isfahan, Isfahan, Iran}
\ead{\mailto{m-baghreri@phys.ui.ac.ir},\
\mailto{rokni@sci.ui.ac.ir},\ \mailto{mhnaderi@phys.ui.ac.ir}}
\begin{abstract}
In this paper we study some basic quantum confinement
effects through investigation of a deformed harmonic oscillator
algebra. We show that spatial confinement effects on a quantum
harmonic oscillator can be represented by a deformation function
within the framework of nonlinear coherent states theory. We
construct the coherent states associated with the spatially
confined quantum harmonic oscillator in a one-dimensional
infinite well and examine some of their quantum statistical
properties, including sub-poissonian statistics and quadrature
squeezing.
\end{abstract}

\maketitle

\section{Introduction}
\indent The harmonic oscillator is one of the models most
extensively used in both classical and quantum mechanics. The
usefulness and simplicity make this model a subject of lots of
studies. One of the most important aspects of quantum harmonic
oscillator (QHO) is its dynamic algebra i.e. Weyl-Heisenberg
algebra. This algebra appears in many areas of modern theoretical
physics, as an example we notice that the one-dimensional quantum
harmonic oscillator was successfully used in second quantization
formalism \cite{field}.\\ \indent Due to the relevance of
Weyl-Heisenberg algebra, some efforts have been devoted to
studying possible deformations of the QHO algebra
\cite{macfarlan}. A deformed algebra is a nontrivial
generalization of a given
 algebra through the introduction of one or more deformation
 parameters, such that, in a certain limit of parameters the
 non-deformed algebra is recovered. A particular deformation of
 Heisenberg algebra has led to the notion of $f$-oscillator \cite{man1}. An
 $f$-oscillator is a non-harmonic system, that from mathematical
 point of view its dynamical variables (creation and
 annihilation operators) are constructed from a non canonical
 transformation through
 \begin{equation}\label{defo}
\hat{A}=\hat{a}f(\hat{n})\hspace
{1cm},\hspace{1cm}\hat{A}^{\dag}=f(\hat{n})\hat{a}^{\dag},
 \end{equation}
where $\hat{a}$ and $\hat{a}^{\dag}$ are the usual (non-deformed)
harmonic oscillator operators with $[\hat{a},\hat{a}^{\dag}]=1$
and $\hat{n}=\hat{a}^{\dag}\hat{a}$. The function $f(\hat{n})$ is
called deformation function which depends on the number of
excitation quanta and some physical parameters. The presence of the 
operator-valued deformation function causes the Heisenberg algebra
of the standard QHO to transform into a deformed Heisenberg
algebra. The nonlinearity in $f$-oscillators means dependence of the
oscillation frequency on the intensity \cite{man2}. On the other
hand, in contrast to the standard QHO, $f$-oscillators have not
equal spaced energy spectrum. For example, if we confine a simple
QHO inside an infinite well, due to the spatial confinement, the
energy levels constitute a spectrum that is not equal spaced.
Therefore, in this case it is
reasonable to investigate the corresponding $f$-oscillator.\\
\indent The confined QHO can be used to describe confinement
effects on physical properties of confined systems. Physical
size and shape of the materials strongly affect
 the nature, dynamics of the electronic excitations, lattice vibrations,
 and dynamics of carriers. For example, in the mesoscopic systems, the
dimension of system is comparable with the coherence length of
carriers and this leads to some new phenomena that they do not
appear in a bulk semiconductor, such as quantum interference
between carrier's motion \cite{mesos}. Recent progress in growth
techniques and development of micromachinig technology in
designing mesoscopic systems and nanostructures, have led to
intensive theoretical \cite{theory} and experimental
investigations \cite{expri} on electronic and optical properties
of those systems. The most important point about the nanoscale
structures is that the quantum confinement effects play the
center-stone role. One can even say, in general, that recent
success in nanofabrication technique have resulted in great
interest in various artificial physical systems (quantum dots,
quantum wires and quantum wells) with new phenomena driven by the
quantum confinement. A number of recent experiments have
demonstrated that isolated semiconductor quantum dots are capable
of emitting light \cite{1}. It becomes possible to combine high-Q
optical microcavities with quantum dot emitters as the active
medium \cite{2}. Furthermore, there are many theoretical attempts
for understanding the optical and electronic properties of
 nanostructures especially semiconductor quantum dots \cite{3}. On the other hand, a nanostructure
 such as quantum dot, is a system that carrier's motion is
 confined inside a small region, and during the interaction with other
 systems, the generated excitations such as phonons, excitons and
 plasmons are confined in small region. In order to describe the physical properties of these
 excitations one can consider them as harmonic oscillator.\\ \indent 
As another application of deformed algebra we can refer to the notion 
of parastatistics \cite{green}. The concept of parastatistics has found 
many application in fractal statistics and anyon theory \cite{wilczek}. In addition 
to the anyon theory, the parastatistics has found many interesting application in supersymmetry 
and non-commutative quantum mechanics \cite{plyu}. \\ \indent The construction of generalized deformed
 oscillators corresponding to well-known potentials and
 study of the correspondence between the properties of the
 conventional potential picture and the algebraic one has been done
 \cite{daska}. Recently, the generalized deformed
 algebra and its associated generalized operators have been considered \cite{mizrahi}.
 By looking at the classical correspondence of the Hamiltonian,
 the potential energy and the effective mass function is obtained. In
 this contribution we derive the generalized operators associated
 with a
 definite potential by comparing the physical properties of system
 and physical results of generalized algebra.\\ \indent One of the most interesting features
of the QHO is the construction of its coherent states as the
eigenfunctions of the annihilation operator. As is well known
\cite{man1}, one can introduce nonlinear coherent states (NLCSs)
or $f$-coherent states as the right-hand eigenstates of the deformed
annihilation operator $\hat{A}$. It has been shown \cite{hame}
that these families of generalized coherent states exhibit various
non-classical properties. Due to these properties and their
applications, generation of these states is a very important issue
in the context of quantum optics. The $f$-coherent states may appear
as stationary states of the center-of-mass motion of a trapped ion
\cite{vogel}. Furthermore, a theoretical scheme for generation of
these states in a coherently pumped micromaser within the
frame-work of intensity-dependent Jaynes-Cummings model has been
proposed \cite{naderi}.\\ \indent One of the most important
questions is the physical meaning of the deformation in the NLCSs
theory. It has
 been shown \cite{nondet} that there is a close connection
between the deformation function appeared in the algebraic
structure of NLCSs and the non-commutative geometry of the
configuration space. Furthermore, it has been shown recently
\cite{mahdi}, that a two-mode QHO confined on the surface of a
sphere, can be interpreted as a single mode deformed oscillator,
whose quantum statistics depends on the curvature of sphere.\\
\indent Motivated by the above-mentioned studies, in the present
contribution we are intended to investigate the spatial
confinement effects on physical properties of a standard QHO. It
will be shown that the spatial confinement leads to deformation of
standard QHO. We consider a QHO confined in a one-dimensional
infinite well without periodic
 boundary conditions, and we find its energy levels, as well as associated ladder
 operators. We show that the ladder operators can be interpreted as a special kind
 of the so-called $f$-deformed creation and annihilation operators \cite{man1}.
  \\ \indent This paper is organized as follows: In section 2, we review some
 physical properties of $f$-oscillator and its coherent states.
  In section 3 we consider the spatially confined QHO in a one-dimensional infinite well and construct
  its associated coherent states. We shall also examine some of their quantum statistical properties,
  including sub-Poissonian statistics and quadrature squeezing. Finally, we summarize our
 conclusions in section 4.
 \section{$f$-oscillator and nonlinear coherent states}
 In this section, we review the basics of the $f$-deformed quantum
 oscillator and the associated coherent states known in the
 literature as nonlinear coherent states. In the first step, to investigate one of the sources of deformation 
  we consider an eigenvalue problem for a given quantum physical system and we focus
 our attention on the properties of
 creation and annihilation operators, that allow to make
 transition
 between the states of discrete spectrum of the system Hamiltonian \cite{wunsch}.
 As usual, we expand the Hamiltonian in its eigenvectors
 \begin{equation}
\hat{H}=\sum_{i=0}^{\infty}E_i|i\rangle\langle i|\:,
\end{equation}
where we have choosed $E_0=0$. We introduce the creation (raising)
and annihilation (lowering) operators as follows
\begin{equation}
\hat{a}^{\dag}=\sum_{i=0}^{\infty}\sqrt{E_{i+1}}|i+1\rangle\langle i|
,\hspace{2cm}\hat{a}=\sum_{i=0}^{\infty}\sqrt{E_{i}}|i-1\rangle\langle
i|\:,
\end{equation}
so that $\hat{a}|0\rangle=0$. These
ladder operators satisfy the following commutation relation
\begin{equation}
[\hat{a},\hat{a}^{\dag}]=\sum_{i=1}^{\infty}(E_{i+1}-E_{i})|i\rangle\langle
i|\:.
\end{equation}
Obviously if the energy spectrum is equally spaced that is, it should be
linear in quantum numbers, as in the case of ordinary QHO, then
$E_{i+1}-E_{i}=c$, where $c$ is a constant and in this condition the commutator of
$\hat{a}$ and $\hat{a}^{\dag}$ becomes a constant (a rescaled
Weyl-Heisenberg algebra). On the other hand, if the energy
spectrum is not equally spaced, the ladder operators of the system
satisfy a deformed Heisenberg algebra, i.e. their commutator
depends on the quantum numbers that appear in the energy spectrum.
This is one of the most
important properties of the quantum $f$-oscillators \cite{man1}.\\
\indent An $f$-oscillator is a non-harmonic system characterized
by a Hamiltonian of the harmonic oscillator form
\begin{equation}\label{hami}
\hat{H}_D=\frac{\Omega}{2}(\hat{A}\hat{A}^{\dag}+\hat{A}^{\dag}\hat{A})\hspace{1cm}
(\hbar=1)\:,
\end{equation}
($\hat{A}=\hat{a}f(\hat{n})$) with a specific frequency $\Omega$
and deformed boson creation and annihilation operators defined in
(\ref{defo}). The deformed operators obey the commutation relation
\begin{equation}\label{comut}
[\hat{A}\:,\:\hat{A}^{\dag}]=(\hat{n}+1)f^2(\hat{n}+1)-\hat{n}f^2(\hat{n})\:.
\end{equation}
 The $f$-deformed Hamiltonian
$\hat{H}_D$ is diagonal on the eingenstates $|n\rangle$ in the
Fock space and its eigenvalues are
\begin{equation}\label{energy}
E_n=\frac{\Omega}{2}[(n+1)f^2(n+1)+nf^2(n)].
\end{equation}
In the limit $f\rightarrow 1$, the ordinary expression $E_n=\Omega
(n+\frac{1}{2})$ and the usual (non-deformed)
commutation relation $[\hat{a}\:,\:\hat{a}^{\dag}]=1$ are recovered. \\
\indent Furthermore, by using the Heisenberg equation of motion
with Hamiltonian (\ref{hami})
\begin{equation}
i\frac{d\hat{A}}{dt}=[\hat{A}\:,\:\hat{H}_D],
\end{equation}
we obtain the following solution for the $f$-deformed operators
$\hat{A}$ and $\hat{A}^{\dag}$
\begin{equation}
\hat{A}(t)=e^{-i\Omega
G(\hat{n})t}\hat{A}(0),\hspace{1cm}\hat{A}^{\dag}(t)=\hat{A}^{\dag}(0)e^{i\Omega
G(\hat{n})t},
\end{equation}
where
\begin{equation}
G(\hat{n})=\frac{1}{2}\left((\hat{n}+2)f^2(\hat{n}+2)-\hat{n}f^2(\hat{n})\right).
\end{equation}
In this sense, the $f$-deformed oscillator can be interpreted as a
nonlinear oscillator whose frequency of vibrations depends
explicitly on its number of excitation quanta \cite{man2}. It is
interesting to point out that recent studies have revealed
strictly physical relationship between the nonlinearity concept
resulting from the $f$-deformation and some nonlinear optical
effects, e.g., Kerr nonlinearity, in the context of atom-field
interaction \cite{naderi1}. \\ \indent The nonlinear
transformation of the creation and annihilation operators leads
naturally to the notion of nonlinear coherent states or
$f$-coherent states. The nonlinear coherent states
$|\alpha\rangle_f$ are defined as the right-hand eigenstates of
the deformed operator
\begin{equation}\label{coh}
\hat{A}|\alpha\rangle_f=\alpha|\alpha\rangle_f\:.
\end{equation}
From Eq.(\ref{coh}) one can obtain an explicit form of the
nonlinear coherent states in a number state representation
\begin{equation}
|\alpha\rangle_f=C\sum_{n=0}^{\infty}\alpha^nd_n|n\rangle,
\end{equation}
where the coefficients $d_n$'s and normalization constant $C$
 are, respectively, given by
\begin{eqnarray}\label{coh1}
d_0&=&1\hspace{0.5cm},\hspace{0.5cm}d_n=\left(\sqrt{n!}[f(n)]!\right)^{-1}\hspace{0.5cm},\hspace{0.5cm}
[f(n)]!=\prod_{j=1}^nf(j),\nonumber \\
C&=&\left(\sum_{n=0}^{\infty} d_n^2|\alpha|^{2n}\right)^{\frac{-1}{2}}.
\end{eqnarray}
In recent years the nonlinear coherent states have been paid much
attentions because they exhibit nonclassical features \cite{hame}
and many quantum optical states, such as squeezed states, phase
states, negative binomial states and photon-added coherent states
can be viewed as a sort of nonlinear coherent states
\cite{hame1}. \\ \indent
\section{Quantum harmonic oscillator in a one dimensional infinite well}
\subsection{$f$-deformed oscillator description of confined QHO}
In this section we consider a quantum harmonic oscillator confined
in a one dimensional infinite well. Many attempts have been done
for solving this problem (see \cite{CQHO},\cite{agu}, and
references therein). In most of those works, authors tried to
solve the problem numerically. But in our consideration we try to
solve the problem analytically, to reveal the relationship between
the confinement effect and given deformation function. We start
from the Schr\"{o}dinger equation ($\hbar=1$)
\begin{equation}\label{shshsh}
\left[-\frac{1}{2m}\frac{d^2}{dx^2}+\frac{1}{2}kx^2+V(x)\right]\psi(x)=E\psi(x),
\end{equation}
where
\begin{displaymath}
V(x)=\left\{\begin{array}{ll} 0 & \textrm{$-a\leq x\leq a$}\\
\infty & \textrm{elsewhere}.
\end{array} \right.
\end{displaymath}
According to the approach introduced in previous section, we can
obtain raising and lowering operators from the spectrum of
Schr\"{o}dinger operator. On the other hand, by comparing the
energy spectrum of particular system with energy spectrum of
general $f$-deformed oscillator (\ref{energy}), one could obtain
deformed raising and lowering operators. Hence, we need an
analytical expression for energy spectrum of the system which
explicitly shows dependence on special quantum numbers. The
original problem, confined QHO (\ref{shshsh}), can be solved only
by using the approximation methods. When applying perturbation
theory, one is usually concern with a small perturbation of an
exactly solvable Hamiltonian system. In the case of confined QHO
we deal with three limits. Inside the well, for small values of
position we have harmonic oscillator, for large values we have an
infinite well and at the positions of the boundaries the two
potentials have the same power. Hence the approximation method can
not lead to acceptable results. Therefore, we model the original
problem by a model potential that has mathematical behavior such
as confined QHO. Instead of solving the Schr\"{o}dinger equation
for the QHO confined between infinite rectangular walls in
positions $\pm a$, we propose to solve the eigenvalue problem for
the potential
\begin{equation}\label{pot}
V(x)=\frac{1}{2}k\left(\frac{\tan(\delta x)}{\delta}\right)^2\:,
\end{equation}
where $\delta=\frac{\pi}{2a}$, is a scaling factor depending on
the width of the well. This potential models a QHO placed in the
center of the rectangular infinite well \cite{zico}. The potential
$V(x)$ (\ref{pot}) fulfills two asymptotic requirements: 1)
$V(x)\rightarrow\frac{1}{2}kx^2$ when $a\rightarrow\infty$ (free
harmonic oscillator limit). 2) $V(x)$ at equilibrium position has
the same curvature as a free QHO,
$\left[\frac{d^2V}{dx^2}\right]_{x=0}=k$. This model potential belongs 
to the exactly solvable trigonometric P\"{o}schl-Teller potentials family \cite{poschl}. 
Stationary coherent states for special kind of this potential have been 
considered \cite{antoine}.\\ \indent Now we
consider the following equation
\begin{equation}
\left[-\frac{1}{2m}\frac{d^2}{dx^2}+\frac{1}{2}k\left(\frac{\tan(\delta
x)}{\delta}\right)^2-E\right]\psi(x)=0\;.
\end{equation}
To solve analytically this equation, we use the factorization
method \cite{fac}. By changing the variable and some mathematical
manipulation, the corresponding energy eigenvalues are found as
\begin{equation}\label{eee}
E_n=\gamma(n+\frac{1}{2})^2+\sqrt{\gamma^2+\omega^2}(n+\frac{1}{2})+\frac{\gamma}{4}\:,
\end{equation}
where $\gamma=\frac{4\pi^2}{32a^2m}$,and
$\omega=\sqrt{\frac{k}{m}}$ is the frequency of the QHO. The first
term in the energy spectrum can be interpreted as the energy of a
free particle in a well, the second term denotes the energy
spectrum of the QHO, and the last term shifts the energy spectrum
by a constant amount. It is evident that if $a\rightarrow\infty$
then $\gamma\rightarrow 0$ and the energy spectrum (\ref{eee})
reduces to the spectrum of a free QHO. As is clear from
(\ref{eee}), different energy levels are not equally spaced.
Hence, confining a free QHO leads to deformation of its dynamical
algebra and we can interpret the parameter $\gamma$ as the
corresponding deformation parameter. In Table \ref{table} the
numerical results associated with the original potential, given in
Ref. \cite{CQHO}, are compared with the generated results from the
model potential under consideration. As is seen, the results are
in a good agreement when boundary size is of order of
characteristic length of the harmonic oscillator. The original
oscillator potential when approaches to the boundaries of the well
 becomes infinite suddenly, while the model potential is smooth and
approaches to the infinity asymptotically. Therefore, the model
potential (\ref{pot}) is more appropriate for the physical
systems. \\ \indent
 If we normalize Eq.(\ref{eee}) to energy quanta of the simple
 harmonic oscillator and introduce the new variables $n+\frac{1}{2}=h$,
$\sqrt{\frac{\gamma^2}{\omega^2}+1}=\eta$, and
$\gamma'=\frac{\gamma}{\omega}$ then it takes the following form
\begin{equation}
E_l=\gamma' h^2+\eta h+\frac{\gamma'}{4}.
\end{equation}
By comparing this spectrum with the energy spectrum of an
$f$-deformed oscillator, given by (\ref{energy}), we find the
corresponding deformation function as
\begin{equation}\label{f1f}
f(\hat{n})=\sqrt{\gamma'\hat{n}+\eta}.
\end{equation}
Furthermore, the ladder operators associated with the confined
oscillator under consideration can be written in terms of the
conventional (non-deformed) operators $\hat{a}$ ,
$\hat{a}^{\dag}$ as follows
\begin{equation}\label{deformo}
\hat{A}=\hat{a}\sqrt{\gamma'\hat{n}+\eta}\hspace{1cm},
\hspace{1cm}\hat{A}^{\dag}=\sqrt{\gamma'\hat{n}+\eta}\,\,\hat{a}^{\dag}.
\end{equation}
These two operators satisfy the following commutation relation
\begin{equation}
[\hat{A},\hat{A}^{\dag}]=\gamma'(2\hat{n}+1)+\eta.
\end{equation}
It is obvious that in the limiting case $a\rightarrow\infty$
($\gamma'\rightarrow 0$,$\eta\rightarrow 1$), the right hand side
of the above commutation relation becomes independent of
$\hat{n}$, and the deformed algebra reduces to a the conventional
Weyl-Heisenberg algebra for a free QHO.\\ \indent Classically,
harmonic oscillator is a particle that attached to an ideal
spring, and can oscillate with specific amplitude. When that
particle be confined, boundaries can affect particle's motion if
the boundaries position be in a smaller distance in comparison
with a characteristic length that particle oscillates within it.
This characteristic length for the QHO is given by
$\frac{\hbar}{m\omega}$ $(\hbar=1)$ , and if
$2a\leq\frac{1}{m\omega}$, then the presence of the boundaries affects
the behavior of QHO, otherwise it behaves like a free QHO.
Therefore, one can interpret $l_0=\frac{1}{m\omega}$ as a scale
length where the deformation effects become relevant.
\subsection{Coherent states of confined oscillator}
\begin{figure}
\begin{center}
\includegraphics[angle=0,width=.5\textwidth]{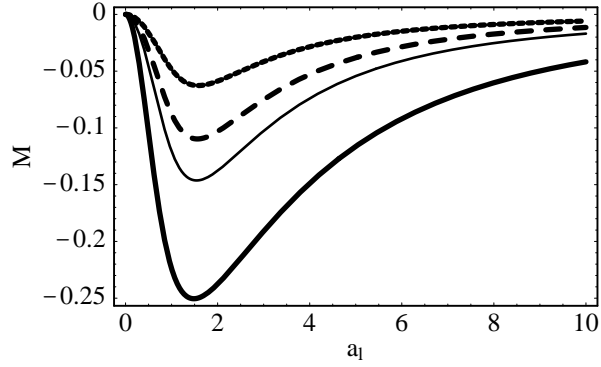}
 \caption{Plots of the Mandel parameter versus the dimensionless parameter
  $a_l=\frac{a}{l_0}$. For $|\beta|^2=0.5$ (dashed curve), for $|\beta|^2=1$ (longdashed curve), for $|\beta|^2=1.5$
  (solid curve) and for $|\beta|^2=4.0$ (bold curve).} \label{f1}
\end{center}
\end{figure}

\begin{figure}
\begin{center}
\includegraphics[angle=0,width=.5\textwidth]{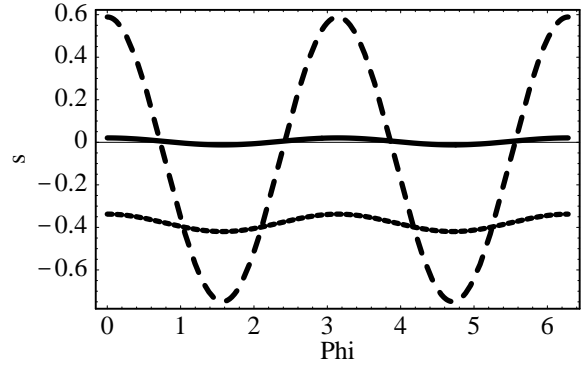}
 \caption{Plot of $s_{\hat X_a}$ versus $\phi$ for $|\beta|^2=4$. The dashed, longdashed and solid curves respectively relate to $a=2.5$,
  $a=1$, $a=0.5$ (the values of $a$ are renormalized to $l_0$).}
\label{f2}
\end{center}
\end{figure}

\begin{figure}
\begin{center}
\includegraphics[angle=0,width=.5\textwidth]{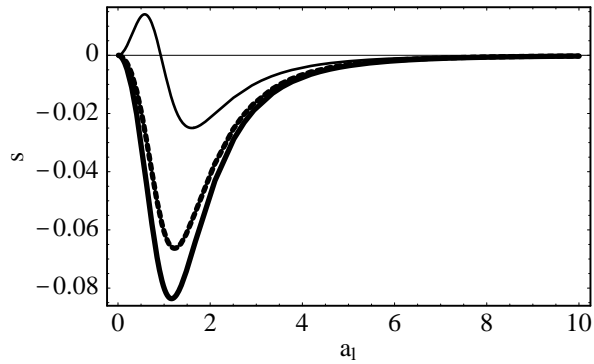}
 \caption{Plots of $s_{\hat X_a}$ versus the dimensionless parameter
  $a_l=\frac{a}{l_0}$ for different phases and $|\beta|^2=1$. Dashed curve, solid curve and bold
  curve
  ,respectively, correspond to $\phi=100$, $\phi=110$ and $\phi=90$.} \label{f3}
\end{center}
\end{figure}

\begin{figure}
\begin{center}
\includegraphics[angle=0,width=.5\textwidth]{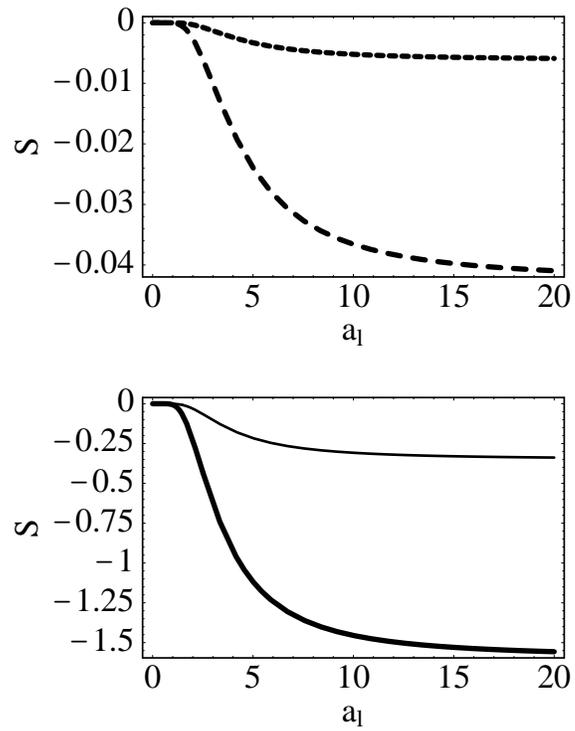}
 \caption{Plots of deformed squeezing parameter $S_{X_A}$ versus the dimensionless parameter
  $a_l=\frac{a}{l_0}$. The dashed curve, longdashed curv, solid curve and bold curve are  respectively, correspond to $|\beta|^2=1$, $|\beta|^2=1.5$ 
   $|\beta|^2=2.5$ and $|\beta|^2=4$.} \label{f4}
\end{center}
\end{figure}

Now, we focus our attention on the coherent states associated
with the QHO under consideration. As usual, we define coherent
states as the right-hand eigenstates of the deformed annihilation
operator
\begin{equation}\label{nlcs}
\hat{A}|\beta\rangle_f=\beta|\beta\rangle_f.
\end{equation}
From (\ref{nlcs}) and using the NLCS formalism introduced in
(\ref{coh})-(\ref{coh1}) the explicit form of the corresponding NLCS of
the confined QHO is written as
\begin{equation}
  |\beta\rangle_f=\mathcal{N}\sum_n\frac{\beta^n}{\sqrt{n!(\gamma' n+\eta)!}}|n\rangle,
\end{equation}
where
$\mathcal{N}=\left(\sum_n\frac{|\beta|^{2n}}{[f(n)!]^2n!}\right)^{-\frac{1}{2}}$
is the normalization factor, $\beta$ is a complex number, and the
deformation function $f(n)$ is given by Eq.(\ref{f1f}). The
ensemble of states $|\beta\rangle_f$ labeled by the single complex
number $\beta$ is called a set of coherent states if the following
conditions are satisfied \cite{klauder}:
\begin{itemize}
  \item normalizability
\begin{equation}\label{c1}
  _f\langle\beta|\beta\rangle_f=1,
\end{equation}
  \item continuity in the label $\beta$
\begin{equation}
  |\beta-\beta'|\rightarrow0
  \hspace{0.5cm}\Rightarrow\hspace{0.5cm}\|\;|\beta\rangle_f-|\beta'\rangle_f\|\rightarrow0,
\end{equation}
  \item resolution of the identity
\begin{equation}\label{mmm}
  \int_c
  d^2\beta|\beta\rangle_{f}{ }_f\langle\beta|w(|\beta|^2)=\hat{I},
\end{equation}
where $w(|\beta|^2)$ is a proper measure that ensures the
completeness and the integration is restricted to the part of the
complex plane where normalization converges.
\end{itemize}
The first two conditions can be proved easily. For the third
condition, we choose the normalization constant as
\begin{equation}
  \mathcal{N}^2=\frac{|\beta|^{\eta}}{I_\eta^{\gamma'}(2|\beta|)},
\end{equation}
where
\begin{equation}
I_\eta^{\gamma'}(x)=\sum_{s=0}^\infty\frac{1}{s!(\gamma'
s+\eta)!}(\frac{x}{2})^{2s+\eta},
\end{equation}
is similar to the modified Bessel function of the first kind of
the order $\eta$ with the series expansion
$I_\eta(x)=\sum_{s=0}^\infty\frac{1}{s!(s+\eta)!}(\frac{x}{2})^{2s+\eta}$.
Resolution of the identity of the deformed coherent states
$|\beta\rangle_f$ can be written as
\begin{eqnarray}
\int
d^2\beta|\beta\rangle_f\langle\beta|w(|\beta|)=&&\pi\sum_n|n\rangle\langle
n |\frac{1}{n!(\gamma' n+\eta)!}\int_0^\infty
d|\beta||\beta||\beta|^{2n}\\ \nonumber
&&\times\frac{|\beta|^\eta}{I_\eta^{\gamma'}(2|\beta|)}w(|\beta|).
\end{eqnarray}
Now we introduce the new variable $|\beta|^2=x$ and the measure
\begin{equation}
  w(\sqrt{x})=\frac{8}{\pi}I_\eta^{\gamma'}(2\sqrt{x})K_m(2\sqrt{x})x^{\frac{l}{2}},
\end{equation}
where $K_m(x)$ is the modified Bessel function of the second kind
of the order $m$, $m=(\gamma'-1)n+\alpha$, and $l=(\gamma'-1)n+1$.
Using the integral relation $\int_0^\infty
K_\nu(t)t^{\mu-1}dt=2^{\mu-2}\Gamma\left(\frac{\mu-\nu}{2}\right)\Gamma\left(\frac{\mu+\nu}{2}\right)$
\cite{bessel}, we obtain
\begin{equation}
\int d^2\beta|\beta\rangle_f
 { }_f\langle\beta|w(|\beta|)=\sum_n|n\rangle\langle n |=\hat{I}.
\end{equation}
 \indent We therefore conclude that the states $|\beta\rangle_f$ qualify as coherent states in the sense
described by the conditions (\ref{c1})-(\ref{mmm}).\\ \indent We
now proceed to examine some nonclassical properties of the
nonlinear coherent states $|\beta\rangle_f$. As an important
quantity, we consider the variance of the number operator
$\hat{n}$. Since for the coherent states the variance of number
operator is equal to its average, deviation from the Poissonian
statistics can be measured with the Mandel parameter \cite{mandel}
\begin{equation}
  M=\frac{(\Delta
  n)^2-\langle\hat{n}\rangle}{\langle\hat{n}\rangle}.
\end{equation}
This parameter vanishes for the Poisson distribution, is positive
for the super-Poissonian distribution (bunching effect), and is
negative for the sub-Poissonian distribution
 (antibunchig effect).
Fig. \ref{f1} shows the size dependence of the Mandel
parameter for different values of $|\beta|^2$. As is seen, the
Mandel parameter exhibits the sub-Poissonian
 statistics and with further increasing values of $a$ it is finally stabilized at an asymptotical zero value
 corresponding to the Poissonian statistics. In addition, the smaller the parameter
 $|\beta|^2$ is, the more rapidly the Mandel parameter tends to the Poissonian statistics.\\ \indent
As another important nonclassical property we examine the
quadrature squeezing. For this purpose we first consider the
conventional quadrature operators $\hat{X}_a$ and $\hat{Y}_a$
defined in terms of nondeformed operators $\hat{a}$ and
$\hat{a}^\dag$ as \cite{scully}
\begin{equation}
  \hat{X}_{a}=\frac{1}{2}(\hat{a}e^{i\phi}+\hat{a}^{\dag}e^{-i\phi})\hspace{1cm}\hat{Y}_a=\frac{1}{2i}(\hat{a}e^{i\phi}
  -\hat{a}^{\dag}e^{-i\phi}).
\end{equation}
In this equation, $\phi$ is the phase of quadrature operators which can effectivly affect the squeezing properties. The commutation relation for $\hat{a}$ and $\hat{a}^{\dag}$ leads
to the following uncertainty relation
\begin{equation}
  (\Delta \hat X_a)^2(\Delta
  \hat Y_a)^2\geq\frac{1}{4}|\langle[\hat{X}_a,\hat{Y}_a]\rangle|^2=\frac{1}{16}.
\end{equation}
For the vacuum state $|0\rangle$, we have $(\Delta \hat
X_a)^2=(\Delta \hat Y_a)^2=\frac{1}{4}$ and hence $(\Delta \hat
X_a)^2(\Delta \hat Y_a)^2=\frac{1}{16}$. A given quantum state of
the QHO is said to be squeezed when the variance of one of the
quadrature components $\hat{X}_a$ and $\hat{Y}_a$ satisfies the
relation
\begin{equation}
  (\Delta \hat O_{a})^2<(\Delta \hat O_{a})^2_{vacuum}=\frac{1}{4}\hspace{0.5cm}  (\hat O_a=\hat X_a\hspace{0.3cm}or\hspace{0.3cm}
  \hat Y_a).
\end{equation}
The degree of quadrature squeezing can be measured by the
squeezing parameter $s_{\hat O}$ defined by
\begin{equation}
   s_{\hat O}=4(\Delta \hat O_a)^2-1.
\end{equation}
Then, the condition for squeezing in the quadrature component can
be simply written as $s_{\hat O}<0$. In Fig. \ref{f2} we have plotted
 the parameter $s_{\hat X_a}$ corresponding to the squeezing of $\hat{X}_a$ with respect to
 the phase angle $\phi$ for three different values of $a$. As is
 seen, the state $|\beta\rangle_f$ exhibits squeezing for different
values of the confinement size, and when $a_l=\frac{a}{l_0}=2.5$, the quadrature
$\hat X_a$ exhibits squeezing for all values of the phase angle
$\phi$. Fig. \ref{f3} shows the plot of $s_{\hat X_a}$ versus the
dimensionless parameter $a_l=\frac{a}{l_0}$ for different values of
the phase $\phi$. As is seen, with the increasing value of
$a_l\;(\frac{a}{l_0})$, the quadrature component tends to the zero according
to the vacuum fluctuation. Let us also consider the deformed
quadrature operators $\hat{X}_A$ and $\hat{Y}_A$ defined in terms
of the deformed operators $\hat{A}$ and $\hat{A}^{\dag}$ as
\begin{equation}
  \hat{X}_{A}=\frac{1}{2}(\hat{A}e^{i\phi}+\hat{A}^{\dag}e^{-i\phi}),\hspace{1cm}\hat{Y}_A=\frac{1}{2i}(\hat{A}e^{i\phi}
  -\hat{A}^{\dag}e^{-i\phi}).
\end{equation}
By considering the commutation relation (\ref{comut}) for the
deformed operators
 $\hat{A}$ and $\hat{A}^{\dag}$, the squeezing condition
 for the deformed quadrature operators $\hat{O}_A$ ($=\hat{X}_A$, $\hat{Y}_A$)can be written
 as
\begin{equation}
S=4(\Delta
\hat O_A)^2-\langle(\hat{n}+1)f^2(\hat{n}+1)\rangle+\langle\hat{n}f^2(\hat{n})\rangle<0.
\end{equation}
In Fig. \ref{f4} we have plotted the parameter $S_{\hat X_A}$
versus the dimensionless parameter $\frac{a}{l_0}$ for three
different values of $|\beta|^2$. As is seen, the deformed
quadrature operator exhibits squeezing for all values of $a$.
Furthermore, with the increasing value of $|\beta|^2$ the
squeezing of the quadrature $\hat X_A$ is enhanced.
\section{Conclusion}
In this paper, we have considered the relation between the spatial
confinement effects and a special kind of $f$-deformed algebra. We
have found that the confined simple harmonic oscillator can be
interpreted as an $f$-oscillator, and we have obtained the
corresponding deformation function. By constructing the associated
NLCSs, we have examined the effects of confinement size on
non-classical statistical properties of those states. The result
show that the stronger confinement leads to the strengthening of
non-classical properties. We hope that our approach may be used in
description of phonons in the strong excitation regimes, photons in a microcavity and
different elementary excitations in confined systems. The work on
this direction is in progress.

\textbf{Acknowledgment} The authors wish to thank
      the Office of Graduate Studies of the University of Isfahan and
      Iranian Nanotechnology initiative for
      their support.
 \linespread{1}
\begin{table}
\begin{center}
\caption{Calculated energy levels of the confined QHO in a one
dimensional infinite well by using our model potential in
comparison with the numerical results given in Ref.\cite{CQHO}}\label{table}
\begin{tabular*}{\textwidth}{@{}l*{15}{@{\extracolsep{0pt plus12pt}}l}}
\br
state& boundary size& model potential& numerical results\\
\hline 0& a=0.5& 4.98495312& 4.95112932\\
0& 1& 1.41089325& 1.29845983\\
0& 2& 0.67745392& 0.53746120\\
0& 3& 0.57321464& 0.50039108\\
0& 4& 0.54003728& 0.50000049\\
\hline 1& a=0.5& 19.88966157& 19.77453417\\
1& 1& 5.46638033& 5.07558201\\
1& 2& 2.34078691& 1.76481643\\
1& 3& 1.85672176& 1.50608152\\
1& 4& 1.69721813& 1.50001461\\
\hline 2& a=0.5& 44.66397441& 44.45207382\\
2& 1& 11.98926850& 11.25882578\\
2& 2& 4.62097017& 3.39978824\\
2& 3& 3.41438455& 2.54112725\\
2& 4& 3.00861155& 2.50020117\\
\hline 3& a=0.5& 79.30789166& 78.99692115\\
3& 1& 20.97955777& 19.89969649\\
3& 2& 7.51800371& 5.58463907\\
3& 3& 5.24620303& 3.66421964\\
3& 4& 4.47421754& 3.50169153\\
\hline
4& a=0.5& 123.82141330& 123.41071050\\
4& 1& 32.43724814& 31.00525450\\
4& 2& 11.03188752& 8.36887442\\
4& 3& 7.35217718& 4.95418047\\
4& 4& 6.09403610& 4.50964099\\
\br
\end{tabular*}
\end{center}
\end{table}
\linespread{1.4}

{}
\newpage
\end{document}